\documentclass[10pt,conference]{IEEEtran}
\usepackage{cite}
\usepackage{amsmath,amssymb,amsfonts}
\usepackage{algorithmic}
\usepackage{graphicx}
\usepackage{textcomp}
\usepackage{xcolor}
\usepackage{xspace}
\usepackage{csquotes}
\usepackage{comment}
\usepackage[colorlinks=true,urlcolor=teal,linkcolor=teal,citecolor=teal]{hyperref}
\usepackage{nicefrac}
\usepackage[inline]{enumitem}
\usepackage{setspace}
\usepackage{booktabs}

\usepackage{mathtools}
\mathtoolsset{showonlyrefs}

\usepackage{siunitx}
\usepackage{subcaption}

\usepackage{tikz}
\usepackage[inline]{enumitem}
\usepackage{listings}
\usepackage[frozencache,cachedir=.]{minted}
\setminted{fontsize=\scriptsize}
\definecolor{LightGray}{gray}{0.95}
\newcommand{\reprourl}{https://github.com/lfd/qce2025-design-automation}
\newcommand{\zenodourl}{https://doi.org/10.5281/zenodo.15209167}

\usepackage{dsfont}

\newcommand{\ket}[1]{| #1 \rangle}

\newcommand{\imag}{\mathrm{i}}

\def\subsectionautorefname{section}

\newcommand{\secref}[1]{%
\begingroup%
\def\chapterautorefname{Chapter}%
\def\sectionautorefname{Section}%
\def\subsectionautorefname{Section}%
\autoref{#1}%
\endgroup%
}

\newcommand{\eg}{\emph{e.g.},\xspace}
\newcommand{\ie}{\emph{i.e.},\xspace}
\newcommand{\etal}{\emph{et al.}\xspace}
\newcommand{\cf}{\emph{cf.}\xspace}
\newcommand{\cnot}{\ensuremath{\text{C\raisebox{0.08em}{--}}\!X}\xspace}

\newcommand{\rz}{R_Z}

\newcommand{\mono}[1]{{\footnotesize \texttt{#1}}}
\newcommand{\smallmono}[1]{{\scriptsize \texttt{#1}}}

\usepackage{censor}
\newboolean{anonymous}
\setboolean{anonymous}{false}
\ifbool{anonymous}{}{\renewcommand{\censor}[1]{#1}}
\ifbool{anonymous}{}{\renewcommand{\blackout}[1]{#1}}
\ifbool{anonymous}{\newcommand{\genemail}[2]{\href{mailto:xxx.xxx@xx.xx}{\blackout{#2}}}}{\newcommand{\genemail}[2]{\href{#1}{#2}}}

\newcommand{\hta}{\censor{High-Tech Agenda Bavaria}}

\begin{document}
\bstctlcite{BSTcontrol}

\title{Predict and Conquer: Navigating Algorithm Trade-offs with Quantum Design Automation}

\author{
    \IEEEauthorblockN{\blackout{Simon Thelen}}
    \IEEEauthorblockA{
        \blackout{\textit{Technical University of}} \\
        \blackout{\textit{Applied Science Regensburg}} \\
        \blackout{Regensburg, Germany} \\
        \genemail{mailto:simon.thelen@othr.de}{simon.thelen@othr.de}
    }
    \and
    \IEEEauthorblockN{\blackout{Wolfgang Mauerer}}
    \IEEEauthorblockA{
        \blackout{\textit{Technical University of}} \\
        \blackout{\textit{Applied Science Regensburg}} \\
        \blackout{\textit{Siemens AG, Technology}} \\
        \blackout{Regensburg/Munich, Germany} \\
        \genemail{mailto:wolfgang.mauerer@othr.de}{wolfgang.mauerer@othr.de}
    }
}

\maketitle

\begin{abstract}
Combining quantum computers with classical compute power has become a standard means for developing algorithms and heuristics that are,
eventually, supposed to beat any purely classical alternatives.
While in-principle advantages for solution quality or runtime are expected for increasingly many approaches, substantial challenges remain:
Non-functional properties like runtime or solution quality of many suggested approaches are not yet fully understood, and need to be explored empirically. This, in turn, makes it unclear which approach is best 
suited for a given problem.
Accurately predicting behaviour and properties of quantum-classical algorithms opens possibilities for software abstraction layers, which in turn can automate decision-making for algorithm selection and parametrisation.
While such techniques find frequent use in classical high-performance computing, they are still mostly absent from quantum software toolchains.

In this paper, we present a methodology (accompanied by a reproducible reference implementation) to perform algorithm selection based on desirable non-functional requirements. This
greatly simplifies decision-making processes for end users.
Based on meta-information annotations at the source code level, our framework traces key characteristics of quantum-classical 
heuristics and algorithms, and uses this information to predict the most suitable approach and its parameters for given computational challenges
and their non-functional requirements.
As combinatorial optimisation is a very extensively studied aspect
of quantum-classical systems, we perform a comprehensive case
study based on numerical simulations of algorithmic approaches to implement and validate
our ideas. We develop statistical models to quantify the influence
of various factors on non-functional properties, and establish
predictions for optimal algorithmic choices without manual 
user effort. We argue that our methodology generalises to problem classes beyond combinatorial optimisation, such as Hamiltonian simulation, and lays a foundation for integrated software layers for quantum design automation.

\end{abstract}


\section{Introduction}
For many practical use cases, quantum computing (QC) has theoretical advantages over classical alternatives regarding execution time or solution quality.
Even in larger, fault-tolerant hardware regimes, due to the specialised nature of many quantum approaches, it is widely accepted that rather than replacing existing systems, quantum computers will likely be integrated into high performance computing (HPC) architectures to accelerate specific tasks such as optimisation or simulation.
This hybrid approach allows HPC (or embedded or specialised) systems to leverage the strengths of both quantum and classical computing.

For many problems classes, end users are faced with a wide choice of quantum algorithms.
For the simulation of physical systems, for instance, various algorithms exist \cite{Suzuki:1991,Berry:2015,Low:2019}, which,
depending on the precise characteristics of the system, have vastly different scaling behaviour, qubit usage and approximation errors.
Finding the optimal technique for a given system is a difficult task.
Similarly, in combinatorial optimisation, many approaches have been proposed \cite{adiabatic,qaoa,wsqaoa,wsinitqaoa,rqaoa,adapt-qaoa,lrqaoa}, including analogue techniques as well as hybrid algorithms like QAOA~\cite{qaoa}.
A recent survey \cite{qaoa-survey} identified 18 different QAOA variants.
Most of them can be further customised through parameters like number of circuit layers or measurements, resulting in a vast array of possibilities.

When integrating quantum algorithms into classical HPC systems, application-specific requirements such as solution quality, execution time or, in the case of non-fault-tolerant hardware, noise resistance must be considered.
This results in a number of trade-offs that need to be carefully assessed.
Choosing a different algorithm or parametrisation can have non-obvious effects on runtime or solution quality.
These effects are further complicated by the interplay of the quantum and classical parts in hybrid algorithms, making it difficult to decide a-priori which algorithm variant performs best for which task.
Ideally, selecting a suitable algorithm and parametrisation depending on application requirements should be handled automatically by the compiler or runtime, as is common in HPC software stacks.
However, the effects of instance and algorithm properties on runtime and solution quality are still insufficiently understood.
As a consequence, design automation techniques are notably missing from most quantum toolchains.

We propose a design automation framework that selects the appropriate quantum algorithm for a given problem instance from a pool of variants based on application-specific non-functional requirements, which are specified by the user in the form of code annotations.
To validate our framework, we consider hybrid classical-quantum algorithms for combinatorial optimisation, since these have been studied extensively in recent years.
Using numerical simulations, we aim to identify factors that influence solution quality and runtime of these algorithms, in both fault-tolerant and noisy hardware regimes.
Our goal is to develop statistical models which predict algorithm behaviour for unknown problem instances based on data collected for a set of baseline instances.
We investigate how such models can be used to perform algorithm selection automatically, given selection criteria provided by the end user at the code level.
We further demonstrate how our proposed framework can be applied to a wider range of problem classes, such as Hamiltonian simulation.
We provide the source code for our framework as well as our models and evaluation results in form of a comprehensive \href{\reprourl}{code repository} and \href{\zenodourl}{reproduction package} (links in PDF) \cite{mauerer-reproduction-package}.

\section{Related Work}
\label{sec:related-work}
Software solutions addressing the challenges of integrating QC into HPC environments are being designed and implemented~\cite{Humble:2021,Alexeev:2024,Karalekas:2020,Bandic:2022,Farooqi:2023,Elsharkawy:2023,Campbell:2023}; we cannot review them all in detail.
Karalekas~\etal~\cite{Karalekas:2020} introduce a framework for a quantum-classical cloud platform, enumerating its architectural requirements and showcasing two platform-level enhancements that optimise the platform for variational hybrid algorithms.
Bandic~\etal~\cite{Bandic:2022} survey QC full-stacks, highlighting the need for tight co-design and vertical integration between software and hardware.
Auto-tuning  in HPC environments was studied by Hoefler~\etal~\cite{Hoefler:2015}.
Wintersperger~\etal~\cite{Wintersperger:2022} and Safi~\etal~\cite{safi:23:codesign} study the influence of parameters like communication latencies and adapted topologies in HPC/QC systems. 
Elsharkawy~\etal~\cite{Elsharkawy:2023} assess the suitability of quantum programming tools for integration with classical HPC frameworks.
Close integration of classical and quantum aspects is a paramount desire in most studies.

In classical HPC, significant efforts have been dedicated to modelling performance in order to optimise runtime, again allowing us to only review a representative sample.
Barnes \etal~\cite{Barnes:2008} explore regression-based approaches to predict parallel program scalability, using execution data from smaller processor counts to forecast performance on larger systems.
Di \etal~\cite{Di2012-xf} demonstrate the effectiveness of Bayesian models for predicting host load in cloud systems, achieving high accuracy compared to traditional methods.
Calotoiu \etal~\cite{Calotoiu2013-oe} introduce automated performance modelling techniques to identify scalability bottlenecks early, enabling developers to address issues before they impact large-scale runs. 

Many hybrid quantum algorithms for combinatorial optimisation have been proposed, with QAOA~\cite{qaoa} being among the most prevalent.
Various similar algorithms have been suggested, which modify circuit structure (\eg Refs.~\cite{Zhang:2017,Wang:2020,Baertschi:2020,rqaoa,Zhu:2022}) or the optimisation process (\eg Refs.~\cite{wsqaoa,wsinitqaoa,Tate:2023,Vijendran:2024,Sud:2024,montanezbarrera:2024,Streif:2020}).
We consider a representative selection for our evaluation~\cite{qaoa,wsqaoa,wsinitqaoa,rqaoa}.

Regarding the simulation of quantum algorithms on noisy hardware, Georgopoulos~\etal~\cite{noise-model} present an approach to simulate effects of three error types using quantum channels, and align the model with experimental observations. Greiwe~\etal~\cite{felix} investigate the effects of imperfections on quantum algorithms. % to show possible options for tailoring NISQ machines to improve system performance in a co-design approach.
Xue~\etal~\cite{noisy-qaoa} confirm the effectiveness of hybrid algorithms on noisy quantum devices by studying effects of quantum noise on standard QAOA. Marshall~\etal~\cite{noisy-qaoa2} provide an approximate model for fidelity and expected cost given noise rate, system size, and circuit depth.

\section{Context and Foundation}\label{sec:foundation}

\subsection{Subject Problems}
To identify predictable patterns in the behaviour of quantum optimisation algorithms, we consider five NP-complete problems: Max-Cut, Minimum Vertex Cover (MVC), Maximum Independent Set (MIS), Partition and Max-3SAT.

Max-Cut is the problem of partitioning the vertices of an undirected graph into two sets $S$ and $T$ such that the number of edges between $S$ and $T$ is minimised.
For MVC, the goal is to find the smallest vertex subset $C$ of an undirected graph such that, for every edge $(u, v)$, $u \in C$ or $v \in C$.
MIS is the complementary problem where the objective is to find the largest vertex subset $C$ such that for no edge $(u, v)$, both $u$ and $v$ are in $C$.
MVC and MIS are closely related since the vertices which are not part of an independent set form a vertex cover and vice versa.

Partition is conceptionally similar to Max-Cut: Here, the objective is to partition a set of numbers into two $S$ and $T$, such that the absolute difference between the sum of numbers in $S$ and the sum of numbers in $T$ is minimised.
Finally, a Max-3SAT problem instance is defined by a set of boolean variables and a set of clauses over these variables.
Each clause is a logical disjunction of exactly three literals where a literal is either a variable or a negated variable.
The objective is to find an assignment of variables that maximises the number of satisfied clauses.

The subject problems
\begin{enumerate*}[label=(\alph*)]
\item are well-understood, with many applications, (\eg Refs.~\cite{DBLP:journals/corr/BianGB16,vertex-cover-application}),
\item can be encoded efficiently for the studied algorithms (\eg~one qubit per vertex for Max-Cut, MIS and MVC) and
\item differ considerably in their hardness of approximation:
For Partition a fully polynomial-time approximation scheme is known~\cite{subset-sum-fptas};
MVC, Max-3SAT and Max-Cut can be approximated in polynomial time within factors of $2$, $0.875$ and $0.878$ respectively\cite{gw-rounding,vertex-cover-approximation,Karloff:2002};
MIS cannot be approximated efficiently within any constant factor unless $\mathrm{P} = \mathrm{NP}$ \cite{Bazgan:2005}.
\end{enumerate*}

\subsection{Subject Algorithms}
\label{sec:qaoa}
We consider four hybrid quantum optimisation algorithms which incorporate classical compute power in different ways.
These algorithms are targetted at problems that can be described as quadratic unconstrained binary optimisation problems (QUBOs).
Many NP-complete optimisation problems, including our subject problems, have efficient QUBO encodings~\cite{ising,Schoenberger:2023,Gogeissl:2024}.
QAOA is one of the most widely studied hybrid quantum optimisation algorithms \cite{qaoa}.
The QAOA circuit first prepares an initial state, typically  $\ket{+}^n$, and applies a series of unitaries: $e^{-\imag \beta_p H_M} e^{-\imag \gamma_p H_C} \dots e^{-\imag \beta_1 H_M} e^{-\imag \gamma_1 H_C}$, where \(p\) chooses the number of \emph{layers}.
$H_C$ is the \emph{problem Hamiltonian}, which encodes the QUBO objective function.
$H_M = \sum_i X^{(i)}$ is called the \emph{mixer Hamiltonian}. 
Optimal parameters $\beta_i, \gamma_i$ ($1 \leq i \leq p$) are found through multiple circuit evaluations using a classical optimiser.
More layers improve, in principle, results at the expense of runtime, but also amplify noise, which decreases solution quality in non-fault-tolerant scenarios.

Warm-starting QAOA variants add another classical component to the standard QAOA approach by first computing an initial guess for the solution classically and then running QAOA to refine this initial solution.
We consider two warm-starting algorithms: WS-Init-QAOA \cite{wsinitqaoa}, which modifies the initial state of the circuit, and WSQAOA \cite{wsqaoa}, which additionally modifies the mixer Hamiltonian based on the initial guess.

Finally, we consider Recursive QAOA (RQAOA)~\cite{rqaoa,rqaoa2}.
This algorithm uses a classical greedy approach to iteratively assign QUBO variables, using QAOA as a subroutine to find the most \enquote{conclusive} QUBO term in each iteration.

\section{Experiments}
\label{sec:experimental-setup}
We evaluate solution quality and runtime of four hybrid quantum optimisation algorithms (QAOA, WSQAOA, WS-Init-QAOA and RQAOA) using ideal and noisy numerical circuit simulations.
We use the \href{https://atos.net/en/solutions/quantum-learning-machine}{Eviden Qaptiva 800} quantum simulation platform and its proprietary software library \emph{QLM} that includes a high-performance, density-matrix-based noisy circuit simulator.
Different numbers of layers ($1 \leq p \leq 7$) are investigated for each variant. 
We evaluate the algorithms using random instances of the five optimisation problems stated above and using well-known QUBO formulations for these problems~\cite{ising,max3sat-qubo}.
For MIS and MVC, the constraint and the objective QUBO are weighted at a ratio of two to one.
For Max-3SAT, we use a QUBO form which requires one qubit per variable and one qubit per clause \cite{max3sat-qubo}.
We consider instances from 5 to 19 qubits.
Since noisy simulations are much more computationally expensive, we limit the number of qubits to 10 and the number of QAOA layers to 4 in the noisy case.
For each problem size (qubit count), we generate 100 random instances.
For Max-Cut, MVC and MIS, random graphs are considered, which are created by inserting an edge between every pair of vertices with probability $0.5$.
Partition instances are generated by drawing numbers uniformly at random from the interval $[0, 1]$.
For each Max-3SAT instance, we select the number of variables uniformly between $1$ and one third times the number of qubits to allow for a range of both easy and hard SAT instances \cite{Krueger:2020}.
For each clause, one of the $(2n)^3$ possible clause configurations (with and without negation) is chosen uniformly.

For all quantum algorithms, SciPy's COBYLA optimiser with a tolerance of \SI{1}{\%} and 150 maximum iterations is used to optimise the circuit parameters~\cite{scipy,cobyla1}. 
The number of circuit measurements per optimiser iteration is fixed at \num{10000}, which is an adequate comprise regarding accuracy and runtime for the studied size regime.
The warm-starting variants use approximate solutions obtained from the Goemans-Williamson algorithm for
Max-Cut~\cite{gw-rounding}, greedy list scheduling for Partition~\cite{list-scheduling1} and a derandomised version of the \enquote{coin-flip algorithm} for Max-3SAT~\cite{Cormen:2022}.
For MVC, we use the classic two-approximation algorithm, which is based on a greedy maximal matching~\cite{Cormen:2022}.
The same algorithm is also used for the MIS warm-starting solutions, using the complement of the solution, since, as stated above, the vertices not part of a vertex cover form an independent set.

\subsection{Noise Model}
\label{sec:noise-model}
\begin{figure*}
    \input{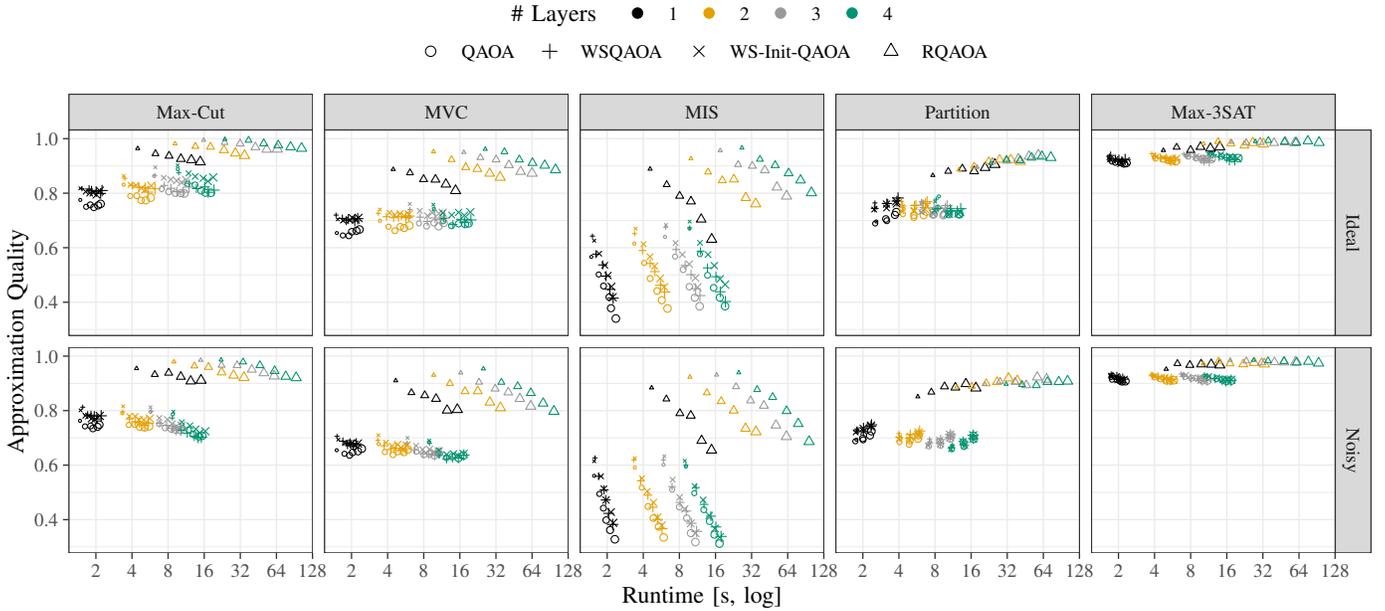}
    \caption{
    Runtime versus approximation quality for ideal and noisy simulations. Each point represents an average over 100 instances for a given number of qubits; point size increases with instance size respectively qubit count (five to ten).
    }
    \label{fig:performance-by-runtime}
\end{figure*}
To ensure the validity of our evaluation for a wide range of fault-tolerant and noisy device types, we consider a continuum of noise levels through density-matrix-based simulations.
We base these simulations of the noise model of Qiskit's noisy simulator, which has seen wide adoption in current QC research~\cite{noise-model,noise-model2,felix}.
It accounts for the dominant sources of noise in most currently available quantum systems and has shown good agreement with experiments on actual quantum hardware \cite{noise-model, noise-model-evaluation,maschek:25:noise}.
Even though the Qiskit model was designed to model the behaviour for IBM superconducting quantum devices, the types of noise it describes generally apply to a broad class of physical qubit realisations.
We implemented the Qiskit model on top of QLM's noisy circuit simulator,
but additionally allow for changing the strength of the individual noise sources cover a wide spectrum of realistic noise regimes.
The noise model considers:
\begin{itemize}
    \item
    \emph{Gate errors:}
    Gate errors are modelled using depolarising channels, which with some depolarising probability $p$ replace the state of the affected qubits by the maximally mixed state \cite{nielsen-chuang}.
    \item
    \emph{Thermal relaxation:}
    Even if the qubit is not involved in any quantum gates, its state slowly transitions to the thermal equilibrium state $\ket 0$ over time, which is modelled as a combination of amplitude and phase damping noise~\cite{nielsen-chuang,superconducting-qubits}.
\end{itemize}
The model is parametrised by
\begin{enumerate*}[label=(\alph*)]
    \item longitudinal (\(T_{1}\)) and transverse (\(T_{2}\)) relaxation time,
    \item gate error and
    duration for each gate type (\eg $\rz$ or $\cnot$),
    \item noise level $l$.
\end{enumerate*}
As a simplifying assumption, we assume all qubits share identical imperfection characteristics.
For a detailed description of our noise model implementation, we refer to our previous work~\cite{thelen:24:noisy-qaoa}.

We transpile from logical gates to the native gate set $\big\{\rz, \sqrt{X}, \cnot\big\}$, as it is supported by many transmon devices.
While not all qubit realisations support $\cnot$ gates natively, equivalents (up to single-qubit rotations) exist on all architectures. 
Since two-qubit gate imperfections usually exceeds single-qubit gate imperfections, any additional single-qubit gates that arise from substituting $\cnot$ gates should not meaningfully affect our results. 

\begin{table}[htbp]
    \centering
    \begin{tabular}{crrrrr}
        \toprule
        & 1-Qubit & 2-Qubit & \(T_{1}\) & \(T_{2}\)\\
        \midrule
        Gate Error &  \SI{0.03}{\percent} & \SI{1}{\percent} \\
        (Gate) Time  & \SI{35}{\nano s} & \SI{400}{\nano s} & \SI{100}{\micro s} & \SI{85}{\micro s} \\
        \bottomrule
    \end{tabular}
\caption{Baseline noise parameters.}\label{tab:noise-parameters}
\end{table}

\secref{tab:noise-parameters} shows baseline noise parameters.
They represent conservative estimates of current transmonic devices.
To study algorithm behaviour on a continuum of noise regimes of today's and future hardware, the parameter $l$ changes the noise level, with $l = 0$ corresponding to noiseless circuits and $l = 1$ corresponding to the baseline parameters shown above.
Specifically, $l$ scales both the error probability $p$ of the depolarising channels and the gate durations.
While our model can easily handle separate noise levels for gate and thermal relaxation noise, our previous work \cite{thelen:24:noisy-qaoa} has shown that these noise sources have a similar effect on circuit fidelity and thus solution quality.
We therefore limit our discussion to only a single dimension covering both noise types.

\subsection{Solution Quality and Runtime Estimation}
Our analysis requires comparable quantities to describe solution quality.
The subject problems include minimisation and maximisation problems, as well as unconstrained and constrained problems (\ie~problems where some solutions are invalid).
To better compare results across problems, we treat them as unconstrained maximisation problems when assigning a value to a solution.
For Max-Cut, MIS and Max-3SAT, the objectives are to maximise cut size, size of the independent set and number of satisfied clauses.
We view Partition as the problem of maximising the size of the smaller set and MVC as the problem of maximizing the reciprocal of the vertex cover size.
For MIS and MVC, we define the value of an invalid solution as the value of the worst-possible solution (one vertex for MIS, all vertices for MVC).
We execute the algorithms using the original QUBO formulations.
Solution quality of an algorithm for an instance is then computed from the circuit's output state, averaged over multiple runs.

Algorithm runtime includes circuit execution, classical parameter optimisation, circuit transpilation and finding initial solution guesses for the warm-starting variants.
To estimate circuit execution time, we use gate durations given in \autoref{tab:noise-parameters}, and a measurement time of \SI{4}{\micro s}.
For classical runtimes, measurements are performed on the simulation hardware.

\subsection{Evaluation Results}
\autoref{fig:performance-by-runtime} compares solution quality and runtime for ideal simulations and simulations at baseline noise levels ($l = 1$).
Each data point represents average approximation quality, that is achieved solution quality divided by the value of the optimal solution, and runtime of an algorithm (QAOA variant and number of layers) for 100 problem instances with the same number of qubits, where we consider instances with 5 to 10 qubits.
We observe that there are clear differences in solution quality and runtime between the studied algorithms.
RQAOA consistently performs best in terms of solution quality, albeit at the expense of runtime, since it executes QAOA multiple times.
As expected, under ideal conditions, both solution quality and runtime increase with increasing number of circuit layers, while under noise there exists a turning point, where the computational power of the added layer does not outweigh the additional noise it introduces.
Solution quality changes at markedly different rates depending on the subject problem, and specific to each algorithm.
However, for each algorithm-problem combination, both solution quality and runtime seem to follow clear and predictable trends.

\section{Solution Quality Models}
We compare multiple approaches to model solution quality, described in detail in the following subsections.
To implement unified models which describe subject problems whose objective functions have vastly different co-domains, our models require two functions $\operatorname{UB}_P(x)$ and $\operatorname{LB}_P(x)$ for each optimisation problem $P$ that map a problem instance to an upper and a lower bound such that the solution quality achieved by an algorithm lies between these two bounds.
As the lower bound, we use the expected value of a solution selected uniformly at random, which any practically-relevant algorithm should be able to beat.
For Max-Cut and Max-3SAT, the expected value of a random solution can be computed analytically.
For other problems where this analytic approach is not feasible, we can approximate the lower bound with good accuracy by randomly sampling the solution space.

For the upper bound, the value of the optimal solution would be a good choice.
However, for the five tested ($\mathrm{NP}$-complete) optimisation problems, computing the optimal value in general is infeasible unless $\mathrm{P} = \mathrm{NP}$.
Instead, we will use problem-specific quantities that are at least as large as the optimal value, but easy to calculate classically, and ideally relatively close to the optimum.
For Max-Cut, Max-3SAT and Partition, we use the trivial upper bounds (cutting all edges, satisfying every clause, distributing the numbers equally).
For MIS, we use the combinatorial upper bound $\nicefrac{1}{2} + \sqrt{\nicefrac{1}{4} + n^2 - n - 2 m}$ \cite{Hansen:1993} where $n$ is the number of vertices and $m$ is the number of edges.
Due to the relationship of MIS and MVC stated above, $n$ minus the MIS bound serves as a bound for MVC.

\subsection{Beta Regression}
For our first approach, we use a Beta regression model \cite{Ferrari:2004}, which is a standard technique to describe continuous response variables lying in a specific interval.
Further, it naturally allows for heteroscedasticity in the response variable, which is present in our data.
We normalize the output variable, namely the expected solution quality $f(x)$, according to
\begin{equation}
\label{eq:normalized-quality}
Y = \frac{f(x) - \operatorname{LB}(x)}{\operatorname{UB}(x) - \operatorname{LB}(x)}
\end{equation}
with $Y \in [0, 1]$.
To account for algorithmic and problem-specific effects, we fit a separate model for each algorithm-problem combination.
Assuming that $Y$ is beta-distributed, using the logit function as the link function and using the problem size (number of qubits) $n$ and the number of QAOA layers $d$ as independent variables, we model $Y$ as $Y \approx \sigma\left(\alpha + \beta n + \gamma d\right)$.
Here, $\sigma$ is the inverse of the logit function: $\sigma(x) = 1 / (1 + \exp(-x))$.
We also considered a version of this model where we apply a Box-Cox transform to the independent variables $n$ and $d$ \cite{Box:1964}.
However, as it achieves similar test errors, we only consider the simpler model without the variable transformation in our evaluation.

\subsection{Power Law Decay}
Our second model is based on the assumption that, for a fixed QAOA variant and a fixed number of layers, normalized solution quality $Y$ decreases with increasing problem size according to a power law.
This model is motivated by visual observation of the behaviour of the QAOA variants as well as the fact that the solution space grows exponentially with the problem size.
To predict solution quality for some QAOA variant with some number of layers for large-scale problem instances, the solution quality for the same variant and the same number must be known for a pool of low-qubit instances.
Let $\overline{Y}_d^{(b)}$ be the average normalized solution quality of some QAOA variant with $d$ layers, according to \eqref{eq:normalized-quality}, for a pool of instances with $b$ qubits.
Then the power law model describes $Y$ as $Y \approx \overline{Y}_d^{(b)} (1 + \alpha (n - b))^{\beta}$ where $n$ denotes the number of qubits.
We find $\alpha$ and $\beta$ using non-linear least squares.
As a variant we also considered an exponential decay model $Y_v \approx \overline{Y}^{(b)}_{d} \exp(-\gamma (n - b))$, where we obtain $\gamma$ using exponential regression.
This, however, achieves sub-par results and is therefore excluded from the analysis.

\begin{figure*}
    \includegraphics[width=\textwidth]{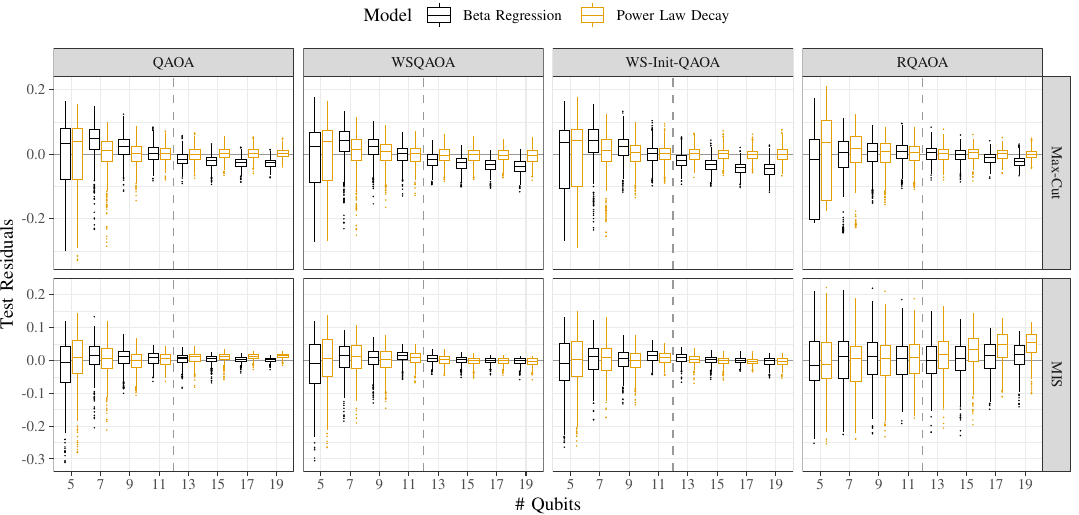}
    \caption{
        Test residuals for models predicting algorithm solution quality on ideal hardware.
        Possible solution values are mapped into $[0, 1]$ for better comparison between problems.
        Training instances range from five to eleven qubits (left of dashed grey line).
        Errors to the right of the line shows extrapolation capabilities of the model.
    }
    \label{fig:ideal-residuals}
\end{figure*}
\subsection{Quality Degradation Model}
The Beta Regression and the Power Law Models can be used to predict the behaviour of the tested quantum algorithms on both ideal and noisy hardware.
This, however, requires fitting separate models for each noise regime in which the algorithm might be performed. 
In heterogenous HPC environments with multiple integrated quantum units with possibly different noise characteristics, one would need to obtain sufficient data on every integrated QPU for each problem and each algorithm to be able to make accurate predictions.
The \emph{Quality Degradation Model} aims to predict how much solution quality degrades, compared to the results on an ideal system, depending on the noise level of the system.

As described in \cite{noisy-qaoa,noisy-qaoa2}, when considering a simplified noise model, which applies one layer of noise 
after every QAOA layer,
solution quality can be approximated as
\begin{equation}
    \label{eq:char-noisy-performance}
    f_\text{noisy}(x) \approx \alpha + (f_\text{ideal}(x) - \alpha) \left(1 - p\beta\right)^{nd}\text{.}
\end{equation}
Here, $f_\text{noisy}$ and $f_\text{ideal}$ denote noisy and ideal QAOA performance, $p$ denotes the error probability, $n$ denotes the number of qubits and $d$ denotes the circuit depth (in the number of QAOA layers); $\alpha$ and $\beta$ are model parameters.
We adapt \eqref{eq:char-noisy-performance} as follows:
It is reasonable to assume that, with increasing circuit depth, the circuit output state approaches the maximally mixed state $I / 2^n$, which is equivalent to randomly guessing a solution, implying that $\alpha = \operatorname{LB}(x)$.
We assume that the noise level $l$ in our model is proportional to the error probability:
$l \propto p$.
This is indeed the case for depolarising noise and also approximately for thermal relaxation noise at small noise levels \cite{thelen:24:noisy-qaoa}.
We replace $p$ with $l$ by absorbing the proportionality coefficient into $\beta$.
In \eqref{eq:char-noisy-performance}, $nd$ denotes the number of noise channels (one per qubit per layer), which we adapt to our more realistic noise model, which includes noise channels after each gate.
To simplify the model and to increase hardware independence due to gate set specifics, we ignore single-qubit gates and focus on the generally much more error-prone $\cnot$ operations.
In particular, we will use the following degradation model:
\begin{equation}
    f_\text{noisy}(x) \approx \operatorname{LB}(x) + (f_\text{ideal} - \operatorname{LB}(x)) (1 - l \beta)^{n d_{\cnot}} (1 - l \gamma)^{n_{\cnot}}\text{.}
\end{equation}
Here, $d_{\cnot}$ is the circuit depth in terms of $\cnot$ gates and $n_{\cnot}$ is the number of $\cnot$ gates in the circuit; $\beta$ and $\gamma$ are model parameters, which we find using non-linear least squares.
The exponent $n d_{\cnot}$ corresponds to the number of thermal relaxation noise channels due to $\cnot$ circuit depth, while $n_{\cnot}$ corresponds to the number of gate error channels.

\subsection{Model Evaluation}
To increase generalisability and reduce training time in deployment, we want our models to learn algorithm behaviour using relatively small problem instances, while still making good decisions on larger, practically relevant instances.
To evaluate the models' extrapolation ability, we reserve the larger half both our noisy and ideal data points in terms of the number of qubits (the \emph{extrapolation set}) for testing.
In the smaller halves (in terms of qubit count), for each optimisation problem and each problem size, we pick 20 of our 100 random instances, which form the \emph{baseline test set}.
The remaining instances are used as training data.

\autoref{fig:ideal-residuals} visualises test data residuals for normalised solution quality $Y$ on ideal hardware as a box plot.
A small spread around $y = 0$ indicates accurate predictions.
The vertical dashed line separates the baseline and extrapolation test sets.
For space and visual clarity, we show only the results for Max-Cut and MIS;
the other results are similar and are included in our \href{\zenodourl}{reproduction package}.
For all tested problem-algorithm combinations, both models achieve a root mean square error below 0.09 on the baseline test set and below 0.07 on the extrapolation set.
In fact, in most cases, the models become more accurate with larger qubit counts.
This effect can be partially attributed to the increasing similarity between instances as the qubit count grows, a consequence of the random instance generation.
In the extrapolation regime, for four of the five tested optimisation problems, one of the two models consistently achieves higher accuracy across all algorithms:
Beta regression yields better results for MVC and Partition; the Power Law Model performs better for Max-Cut and Max-3SAT.
The Power Law Model also achieves slightly lower error rates for MIS.
In general, prediction errors are highest for RQAOA.

\begin{figure*}
    \centering
    \input{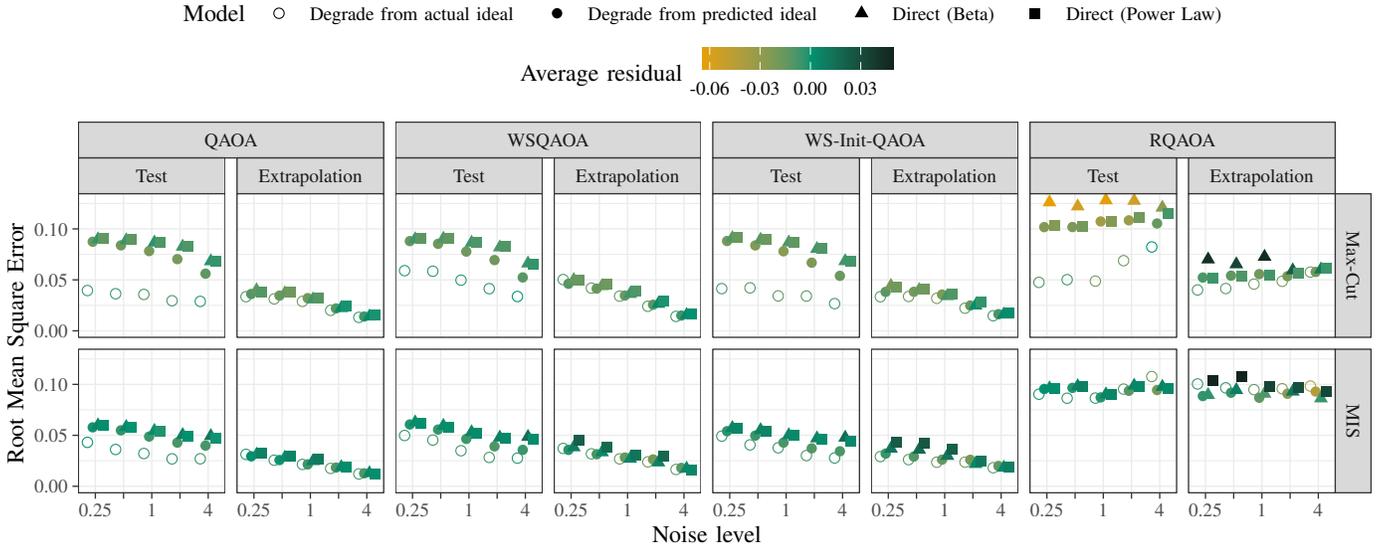}
    \caption{
        Predicting solution quality in noisy environments:
        Test error for different models and noise levels (level 1 approximates current superconducting systems).
        Models were evaluated with instances of the same size as the training instances ($5$ to $7$ qubits) and with larger instances ($8$ to $10$ qubits).
        Yellow points indicate that the model generally underestimates solution quality, black points indicate overestimation.
        Green points indicate no structural over- or underestimation.
    }
    \label{fig:degradation-vs-direct}
\end{figure*}
In the noisy regime, we can either predict solution quality directly or use the Quality Degradation Model to predict how much solution quality decreases, compared to fault-tolerant hardware.
\autoref{fig:degradation-vs-direct} compares the root mean square error of both approaches for noise levels $l \in \{0.25, 0.5, 1, 2, 4\}$.
Point colour visualises mean residuals, where yellow indicates principal quality underestimation and black indicates overestimation.
Again, for visual clarity, we show only the results for Max-Cut and MIS, as the the other results are similar.
For the Quality Degradation Model, we consider the case where ideal solution quality is known and the case where ideal quality is predicted using one of the other models.
For the latter case, we use the better model for the corresponding problem, according to the data from \autoref{fig:ideal-residuals}.
All tested approaches show good model accuracy for the studied problems, algorithms and noise levels.
Among the direct models, the model achieving better results in the ideal case also mostly performs better in the noisy case.
Predicting noisy solution quality from known ideal quality achieves the lowest errors in many cases.
Predicting noisy solution quality by first predicting ideal quality and then predicting quality degradation achieves similar errors compared to predicting noisy quality directly.
This is notable since the direct models need to be trained specifically for each noise level, while the Quality Degradation Model works for any noise level by including noise as a co-variable.

\begin{figure}
    \centering
    \input{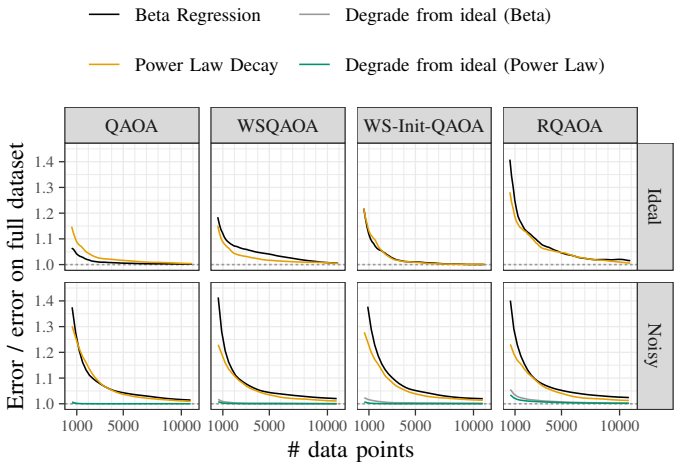}
    \caption{
        Solution quality model error convergence with increasing training set size.
        Root mean square error (extrapolation regime) for partial data set (baseline regime) divided by error for complete baseline set.
    }
    \label{fig:learning-curve}
\end{figure}
Using statistical models to select the best algorithm is only feasible if the number of training problem instances required for good model accuracy is not too large.
We therefore compare extrapolation test error for training sets of different sizes, which we obtain by sampling from the baseline instances.
\autoref{fig:learning-curve} shows the average error of models trained on a subset of the baseline instances, divided by the error achieved after being trained on the entire baseline set.
As the number of instances and thus the number of data points increases, the error approaches the optimal error achieved with the full data sets, which include \num{14000} data points (ideal) and \num{30000} data points (noisy) respectively.
For all algorithm-problem combinations and in both ideal and noisy regimes, 3000 data points or less than 100 instances per problem suffice to be \SI{10}{\percent} off the optimal error.
This indicates that results for a relatively small pool of sufficiently diverse baseline instances can be enough to make well-founded automatic algorithm selections, assuming that the behaviour of algorithms does not fundamentally change for larger instances.

\section{Runtime Model}
\begin{figure}
    \centering
    \includegraphics[width=\columnwidth]{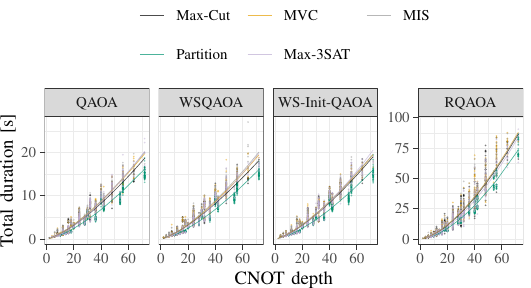}
    \caption{Noisy algorithm runtime, depending on circuit depth (in $\cnot$ gates): polynomial fit. 500 random data points are shown for each problem-algorithm combination.}
    \label{fig:noisy-runtime}
\end{figure}

The runtime of an algorithm depends on factors such as circuit depth, number of shots, number of parameter optimiser iterations (for variational algorithms), transpilation times or communication latencies.
Many factors have a linear effect on algorithm runtime, making it generally easier to model runtime than solution quality.
In our simulations, circuit execution was by far the most runtime-intensive part.
As $\cnot$ gates generally take much longer to execute than single-qubit gates,
the circuit execution time approximately linearly depends on $\cnot$ circuit depth.
For the tested algorithms, runtime is not necessarily linear in circuit depth as it also depends on the optimiser iteration count (which is not fixed).
RQAOA additionally performs multiple QAOA runs on circuits, which are not known a-priori as they depend on intermediate results.
For our framework, we use a relatively simple but flexible approach where we model the runtime $T(x)$ as $T(x) \approx \alpha d_{\cnot}^\beta$.
Here, $d_{\cnot}$ is the circuit depth of the QAOA circuit (the first circuit in the case of RQAOA) in $\cnot$ gates. 
We fit the model using linear least squares after applying a $\log$-transformation.
The fit of the first model for results at noise level $l = 1$ is visualised in \autoref{fig:noisy-runtime}.
The data for other noise levels and the ideal simulations are similar.
By incorporating more model variables besides circuit depth and by specifically tailoring the model to the specific scaling behaviour of the quantum algorithm, a more accurate model can be achieved.
However, since we already achieve good accuracy with such a simple model, we stay with it for the purposes of our framework in order to increase the generalisability of our approach.

\section{Software Framework}
\begin{table}[]
    \centering
    \caption{Variables supported by the algorithm selection framework to specify selection constraints and preferences.}
    \label{tab:framework-variables}
    \begin{tabular}{ll}
        \toprule
        Variable & Semantics\\
        \midrule
        \smallmono{RUNTIME} & $T(x)$\\
        \smallmono{SOLUTION\_QUALITY} & $f(x)$ \\
        \smallmono{RELATIVE\_SOLUTION\_QUALITY} & $\nicefrac{f(x)}{\operatorname{UB}(x)}$ \\
        \smallmono{SOLUTION\_QUALITY\_PER\_RUNTIME} & $\nicefrac{f(x)}{T(x)}$ \\
        \smallmono{RELATIVE\_SOLUTION\_QUALITY\_PER\_RUNTIME} & $\nicefrac{f(x)}{(\operatorname{UB}(x) T(x))}$ \\
        \smallmono{RUNTIME\_PER\_SOLUTION\_QUALITY} & $\nicefrac{T(x)}{f(x)}$ \\
        \smallmono{RUNTIME\_PER\_RELATIVE\_SOLUTION\_QUALITY} & $\nicefrac{T(x)}{(\operatorname{UB}(x) f(x))}$ \\
        \bottomrule
    \end{tabular}
\end{table}
\begin{figure*}
    \centering
    \includegraphics[width=\textwidth]{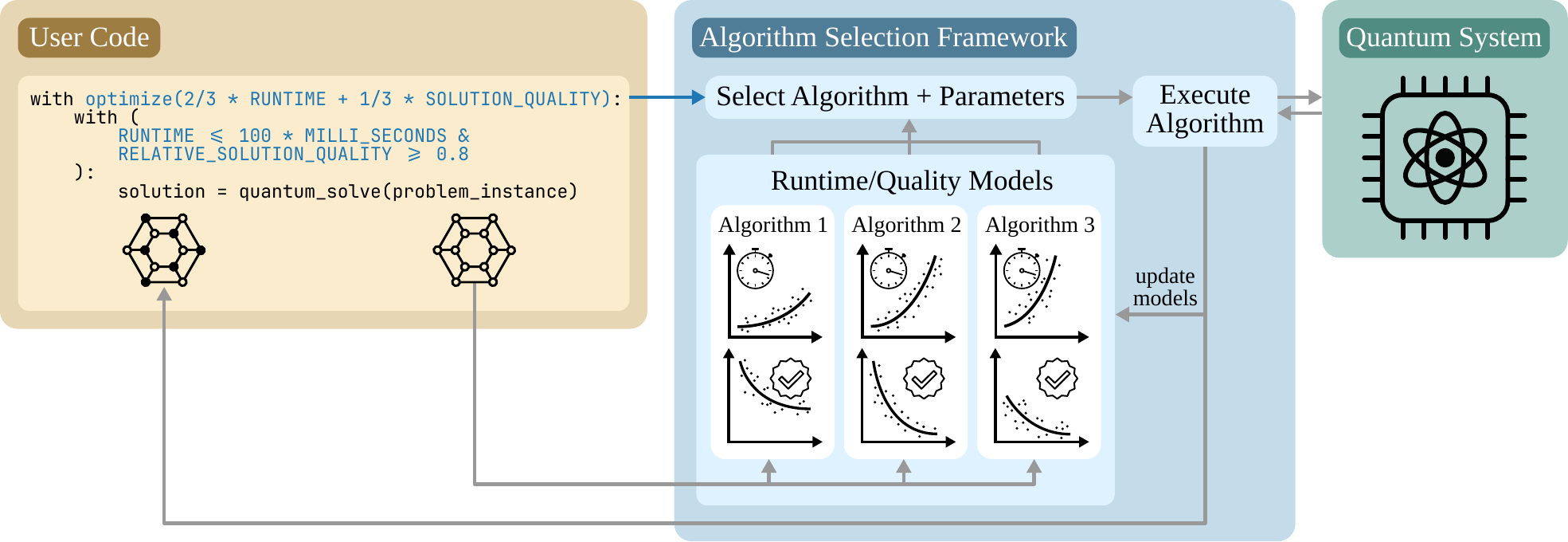}
    \caption{Outline of the structure of the proposed software framework. Given a problem instance, solution quality and runtime are predicted using regression models. Depending on application requirements specified via code annotations, the best algorithm is selected and executed. Runtime and solution quality achieved by the algorithm is used to update the models.}
    \label{fig:architecture}
\end{figure*}
As demonstrated above, in many cases, solution quality and runtime of the algorithms can be predicted using easy to compute quantities such as circuit depth or the expected quality of a random solution.
In order to help engineers integrate quantum algorithms into HPC environments as efficiently as possible, we propose a software framework, based on these results, that can automatically select the optimal algorithm for a given problem instance based on multiple constraints.
The proposed framework uses Python context managers to provide an intuitive but flexible interface through Python's \texttt{with} blocks that can be integrated easily into typical controls structures such as loops without needing to change the specification of the language.
We used Python since it is the most widely used programming language for QC, largely due to the popularity of the \emph{Qiskit} software framework.
In the context of HPC, compiled languages with manual memory management like C++ are of course much more prevalent as they provide significant performance advantages.
Similar syntax constructs to the ones we propose can be realised in C++ by using \texttt{\#pragma} directives.

The provided framework serves as a proof-of-concept to demonstrate that the integration of quantum algorithm selection into existing software stacks is achievable in practice.
\autoref{fig:architecture} visualises the main structure of the framework.
The user passes the problem instance they want to solve to the framework via a function call.
The framework maintains a database which stores solution quality and runtime achieved by the available algorithms for the previously submitted problem instances.
It also stores runtime and solution quality models for each algorithm, which it regularly updates based on recently obtained instance-algorithm pairs.
Based on our model results, the framework uses the model with the lowest error for a given problem.
\begin{figure}
        \begin{minted}[bgcolor=LightGray,mathescape,autogobble]{python}
        from quantum_framework import *
        import networkx as nx

        # load benchmark instances, train model
        framework = AlgorithmSelectionFramework()
        framework.add_results(load_results())

        instance = MaxCut(nx.erdos_renyi_graph(n = 30, p = 0.5))
        
        # find sufficiently large cuts quickly 
        with minimize(RUNTIME):
            with RELATIVE_SOLUTION_QUALITY >= 0.75:
                result = framework.quantum_solve(instance)
            with SOLUTION_QUALITY >= 200:
                result = framework.quantum_solve(instance)

        # favor quality over runtime (ratio 2:1)
        with maximize(2/3 * SOLUTION_QUALITY - 1/3 * RUNTIME):
            result = framework.quantum_solve(instance)

        # find best solution in 10s, given there is a 10%
        # improvement per second
        with maximize(SOLUTION_QUALITY), (
            (RUNTIME <= 10 * SECONDS) &
            (RELATIVE_SOLUTION_QUALITY_PER_RUNTIME >= 0.1)
        ):
            result = framework.quantum_solve(instance)
        \end{minted}
    \caption{
        Code example showcasing our framework from the user perspective; algorithm selection considers requirements and preferences regarding the runtime-quality trade-off, specified using \texttt{with} blocks.
    }
    \label{fig:framework-listing}
\end{figure}
The user describes application requirements by specifying multiple constraints and whether the framework should prioritise runtime, quality or a combination of the two when selecting the appropriate algorithm.
Constraints and selection objectives include variables such as runtime, solution quality or \enquote{quality efficiency}, that is solution quality per runtime (\cf~\autoref{tab:framework-variables}).
For each instance submitted, the framework finds the algorithm that maximises or minimises one of these variables (or a linear combination of several variables) under linear constraints, also defined in terms of these variables.
This allows the user to express nuanced requirements such as \enquote{Maximise solution quality but only if the increased quality is worth the extra computational effort}, while keeping the interface intuitive and easy to understand.
Selection objectives and constraints are specified in code using Python \mono{with} statements, leading to concise and legible code.
It also allows the user to provide constraints and objectives to multiple quantum calls using the scope of the \mono{with} block.
Examples of the proposed syntax and three possible application scenarios are shown in \autoref{fig:framework-listing}.
The code and further examples are provided as part of the \href{\zenodourl}{reproduction package}.

\section{Use Case Illustration:\texorpdfstring{\\}{ }Hamiltonian Simulation}
We demonstrated how QC-HPC integration can be simplified by automating algorithm selection, based on non-functional requirements, specified via suitable code abstraction and annotation mechanisms.
Although we focused on combinatorial optimisation problems, we believe that the main insights of our study to extend far beyond the current NISQ (noisy intermediate-scale quantum) era and are applicable to a wider class of problems on both noisy and fault-tolerant hardware.
To demonstrate the generalisability of our framework, we discuss \emph{Hamiltonian simulation} as an additional use case.
Hamiltonian simulation has broad applications in physics and chemistry and is likely to be one of the first practical use cases of QC~\cite{Georgescu:2014}.
Given a Hamiltonian of a physical system $\hat{H}$, the goal is to simulate a unitary $U$, which approximates the time evolution $e^{-\imag \hat{H} t}$ for a given time $t$, that is, $||U - e^{-\imag \hat{H} t}|| \leq \epsilon$ with approximation quality $\epsilon$.
Similar to combinatorial optimisation, many algorithmic approaches with different parametrisations exist to tackle this problem \cite{Georgescu:2014}, with different approaches showing different scaling behaviour regarding approximation quality and runtime, depending on $t$ and properties of $\hat{H}$.

For our discussion, we consider two scenarios where the framework proposed in this work could aid in choosing the optimal algorithm and parametrisation:
For the Trotter-Suzuki algorithm \cite{Suzuki:1991}, $\hat{H}$ is decomposed into a sum of terms $\hat{H} = \sum_i \hat{H}_i$ such that the time evolution of each $\hat{H}_i$ can be efficiently simulated on a (gate-based) quantum computer.
The time evolution of $\hat{H}$ can then be approximated as $(\prod_i e^{-\imag \hat{H}_i t / d})^d$.
This method experiences a similar time-quality trade-off to the tested QAOA variants (approximation quality and runtime increases layer count $d$).
In fact, QAOA can be interpreted as the Trotterised time evolution of the time-dependent Hamiltonian $H(t) = (1-t) H_M + t H_P$, so it seems very likely that statistical approaches like the one proposed in this work can lead to accurate predictions for the Trotter-Suzuki algorithm for both fault-tolerant and noisy regimes.

\begin{figure}
    \begin{minted}[bgcolor=LightGray,mathescape,autogobble]{python}
        with maximize(INV_SIMULATION_ERROR_PER_RUNTIME), (
            SIMULATION_ERROR <= 0.001
        ):
            hamiltonian = Hamiltonain(...)
            result = framework.evolve_hamiltonian(
                hamiltonian,
                psi_0 = "0" * n, observable="Z" * n,
            )
        \end{minted}
    \caption{Code example demonstrating how the proposed framework can be extended support the additional use case of Hamiltonian simulation (opposed to quantum optimisation).}
    \label{fig:simulation-listing}
\end{figure}
For the second scenario, we consider analogue QC, which has shown great potential in quantum simulation tasks~\cite{Hangleiter:2022,Daley:2022}.
Hamiltonians with specific properties can be decomposed into their core building blocks such that each component has a direct counterpart on the analogue quantum system, allowing for a very direct simulation of the quantum system of interest.
Assuming the existence of fault-tolerant quantum hardware, the Trotter-Suzuki algorithm provides a flexible procedure to simulate a wider class of Hamiltonians with the ability to precisely control the time-quality trade-off, at the expense of runtime due to error correction overheads.
Hybrid approaches also exist that combine both analogue and digital or gate-based techniques \cite{Lamata:2018}. 
Choosing the right algorithm or combination of algorithms again is a highly non-trivial problem as it depends on the properties of the system, the tolerated error and possible runtime constraints.
This task should ideally be abstracted away and performed automatically by the toolchain or software stack.
\autoref{fig:simulation-listing} shows how this automatic selection could look like in our proposed software framework.

\section{Threats to Validity}
\label{sec:threats}

\subsection{Circuit Connectivity}
To simplify our analysis, we assumed that 2-qubit gates can be applied to any pair of qubits, ignoring connectivity constraints present in many hardware realisations.
Qubit routing techniques to handle limited connectivity increase circuit runtime and, on noisy hardware, decrease solution quality due to additional gate errors.
Optimal qubit routing is a hard problem and an active topic of research~\cite{qubit-routing1,qubit-routing2,qubit-routing3}.
However, worst case bounds are known for some classes of problems \cite{Hagge:2020,Weidenfeller:2022}.
Recent results also show that the number of required swap gates follows predictable trends and decreases rapidly with growing qubit connectivity \cite{safi:23:codesign}.
Since our models work well for larger noise levels due to additional swap gates (cf.~\autoref{fig:degradation-vs-direct}), we believe that our approach can be extended to account for settings with limited connectivity.

\subsection{Problem Instance Selection}
We only considered random problem instances in our evaluation.
Since our models achieve good accuracy for a diverse set of problems, we believe our approach to apply to other classes of instances.
However, unevenly distributed instances could impair model accuracy.
This could be handled by identifying other instance properties with a strong effect on solution quality and runtime.
As solution quality correlates with problem hardness in our experiments (\eg Max-Cut is easier than MIS for both classical and quantum techniques), indicators for classically hard instances \cite{Mertens:2003,Dunning:2018,Krueger:2020} could also predict quantum solution quality.
Although we specifically evaluated the ability of the models to generalise to larger qubit counts, we only considered instances with up to 19 qubits.
It is reasonable to hypothesise that the general trends observed persist in practically relevant size regimes, although this, of course, cannot be guaranteed with current hardware~\cite{McGeoch2023:}.

\subsection{Completeness and Confounders}
In our evaluation, we focused on algorithm variant, circuit depth, gate count, qubit count as the main co-factors of our models.
While other factors (like number of measurements, choice of optimiser, number of  optimiser iterations, \dots)  can also lead to quality-runtime trade-offs, we used universally accepted default choices (\num{10000} measurement shots for each
configuration; COBYLA with tolerance $10^{-2}$ and at most 150 maximum).
We varied these factors on a subset of our experimental setup (see the \href{\zenodourl}{reproduction package} for detailled results), and found that while there is limited impact on solution quality and runtime, any general trends that we 
have identified for the standard co-variables appear to stay consistent.
Yet, we did not perform a complete statistical a-priori
analysis to systematically identify and quantify possible confounding factors, or interaction effects between 
co-variables. A substantially more complete statistical approach would be required, which goes beyond the scope of this paper.

\section{Discussion \& Conclusion}
\label{sec:conclusion}

We presented a novel approach for automatically selecting quantum algorithms based on systematically and quantitatively specified application requirements using statistical models.
We validated the effectiveness of our ideas using hybrid algorithms for solving hard optimisation problems as a potential use case.
Our results show that problem-specific and algorithm-specific factors have predictable effects on solution quality and runtime.
In our experiments, these trends continue for instances with up to \SI{75}{\percent} higher qubit counts as were used to train the predictors,
allowing our models to achieve good prediction accuracy from just one hundred baseline instances.
We introduced a reproducible software framework to select the 
best algorithm for given desiderata, and discussed how our approach can be applied to other problem domains such as Hamiltonian simulation.

Given the numerous limitations of quantum hardware and software, realistic estimates of what kinds of results end users can expect from existing algorithms when executed on available quantum hardware are essential for the successful integration QC into HPC applications.
Our results indicate that in many scenarios, relatively few baseline instances suffice to make reliable estimates on runtime and solution quality for previously unseen instances.
Regarding solution quality, we have demonstrated that it is possible to successfully separate algorithm-specific from hardware-specific effects by showing that a unified model predicting solution quality degradation due to hardware failures leads to similar prediction errors as using models specifically trained for a particular noise regime.
Nonetheless, further work is required
to model additional relevant influence factors, and to confirm the applicability of our approach in additional scenarios and problem domains.

Our findings may be particularly useful in the context of rapidly advancing quantum hardware, as they may allow for predictions to be made about how much improvements one can expect from hardware with higher error resistance.

Our work contributes to a vision where compilers or runtime systems can select the most appropriate algorithm based on user-specified requirements, which can significantly reduce efforts and expertise required to solve complex problems using QC, enabling widespread adoption in various fields.
We believe that a wide range of design decisions can be automated using our approach, both within and beyond the field of combinatorial optimisation.

\newcommand{\WM}{\censor{WM}\xspace}
\small{
\textbf{Acknowledgements}
This work is supported by \blackout{the German Federal Ministry of Education and Research within the funding program \emph{Quantentechnologien -- von den Grundlagen zum Markt}}, contract number \censor{13N16092}. \WM acknowledges support by the \hta.
}

\newpage\bibliographystyle{IEEEtran}
\bibliography{references}

% Generated by IEEEtran.bst, version: 1.14 (2015/08/26)
\begin{thebibliography}{10}
\providecommand{\url}[1]{#1}
\csname url@samestyle\endcsname
\providecommand{\newblock}{\relax}
\providecommand{\bibinfo}[2]{#2}
\providecommand{\BIBentrySTDinterwordspacing}{\spaceskip=0pt\relax}
\providecommand{\BIBentryALTinterwordstretchfactor}{4}
\providecommand{\BIBentryALTinterwordspacing}{\spaceskip=\fontdimen2\font plus
\BIBentryALTinterwordstretchfactor\fontdimen3\font minus \fontdimen4\font\relax}
\providecommand{\BIBforeignlanguage}[2]{{%
\expandafter\ifx\csname l@#1\endcsname\relax
\typeout{** WARNING: IEEEtran.bst: No hyphenation pattern has been}%
\typeout{** loaded for the language `#1'. Using the pattern for}%
\typeout{** the default language instead.}%
\else
\language=\csname l@#1\endcsname
\fi
#2}}
\providecommand{\BIBdecl}{\relax}
\BIBdecl

\bibitem{Suzuki:1991}
M.~Suzuki, ``\BIBforeignlanguage{en}{General theory of fractal path integrals with applications to many-body theories and statistical physics},'' \emph{\BIBforeignlanguage{en}{J. Math. Phys.}}, vol.~32, no.~2, pp. 400--407, Feb. 1991.

\bibitem{Berry:2015}
D.~W. Berry, A.~M. Childs, R.~Cleve \emph{et~al.}, ``\BIBforeignlanguage{en}{Simulating hamiltonian dynamics with a truncated taylor series},'' \emph{\BIBforeignlanguage{en}{Phys. Rev. Lett.}}, vol. 114, no.~9, p. 090502, Mar. 2015.

\bibitem{Low:2019}
G.~H. Low and I.~L. Chuang, ``\BIBforeignlanguage{en}{Hamiltonian simulation by qubitization},'' \emph{\BIBforeignlanguage{en}{Quantum}}, vol.~3, no. 163, p. 163, Jul. 2019.

\bibitem{adiabatic}
\BIBentryALTinterwordspacing
T.~Albash and D.~A. Lidar, ``Adiabatic quantum computation,'' \emph{Reviews of Modern Physics}, vol.~90, no.~1, jan 2018. [Online]. Available: \url{https://doi.org/10.1103%2Frevmodphys.90.015002}
\BIBentrySTDinterwordspacing

\bibitem{qaoa}
E.~Farhi, J.~Goldstone, and S.~Gutmann, ``A quantum approximate optimization algorithm,'' 2014.

\bibitem{wsqaoa}
\BIBentryALTinterwordspacing
D.~J. Egger, J.~Mare{\v{c}}ek, and S.~Woerner, ``Warm-starting quantum optimization,'' \emph{Quantum}, vol.~5, p. 479, jun 2021. [Online]. Available: \url{https://doi.org/10.22331\%2Fq-2021-06-17-479}
\BIBentrySTDinterwordspacing

\bibitem{wsinitqaoa}
A.~Awasthi, F.~B{\"a}r, J.~Doetsch \emph{et~al.}, ``Quantum computing techniques for multi-knapsack problems,'' in \emph{Intelligent Computing}.\hskip 1em plus 0.5em minus 0.4em\relax Springer Nature, 2023.

\bibitem{rqaoa}
S.~Bravyi, A.~Kliesch, R.~Koenig \emph{et~al.}, ``Obstacles to variational quantum optimization from symmetry protection,'' \emph{Phys. Rev. Lett.}, vol. 125, 12 2020.

\bibitem{adapt-qaoa}
L.~Zhu, H.~L. Tang, G.~S. Barron \emph{et~al.}, ``An adaptive quantum approximate optimization algorithm for solving combinatorial problems on a quantum computer,'' 2022.

\bibitem{lrqaoa}
J.~A. Montanez-Barrera and K.~Michielsen, ``Towards a universal {QAOA} protocol: Evidence of a scaling advantage in solving some combinatorial optimization problems,'' 2024.

\bibitem{qaoa-survey}
\BIBentryALTinterwordspacing
K.~Blekos, D.~Brand, A.~Ceschini \emph{et~al.}, ``A {Review} on {Quantum} {Approximate} {Optimization} {Algorithm} and its {Variants},'' Jun. 2023, arXiv:2306.09198 [quant-ph]. [Online]. Available: \url{http://arxiv.org/abs/2306.09198}
\BIBentrySTDinterwordspacing

\bibitem{mauerer-reproduction-package}
W.~Mauerer and S.~Scherzinger, ``1-2-3 reproducibility for quantum software experiments,'' \emph{IEEE SANER}, pp. 1247--1248, 2022.

\bibitem{Humble:2021}
T.~S. Humble, A.~McCaskey, D.~I. Lyakh \emph{et~al.}, ``Quantum computers for high-performance computing,'' \emph{IEEE Micro}, vol.~41, no.~5, pp. 15--23, 2021.

\bibitem{Alexeev:2024}
\BIBentryALTinterwordspacing
Y.~Alexeev, M.~Amsler, M.~A. Barroca \emph{et~al.}, ``Quantum-centric supercomputing for materials science: A perspective on challenges and future directions,'' \emph{Future Generation Computer Systems}, vol. 160, pp. 666--710, 2024. [Online]. Available: \url{https://www.sciencedirect.com/science/article/pii/S0167739X24002012}
\BIBentrySTDinterwordspacing

\bibitem{Karalekas:2020}
\BIBentryALTinterwordspacing
P.~J. Karalekas, N.~A. Tezak, E.~C. Peterson \emph{et~al.}, ``A quantum-classical cloud platform optimized for variational hybrid algorithms,'' \emph{Q.\ Sci.\ \& Tech.}, vol.~5, no.~2, mar 2020. [Online]. Available: \url{https://dx.doi.org/10.1088/2058-9565/ab7559}
\BIBentrySTDinterwordspacing

\bibitem{Bandic:2022}
M.~Bandic, S.~Feld, and C.~G. Almudever, ``Full-stack quantum computing systems in the {NISQ} era: algorithm-driven and hardware-aware compilation techniques,'' in \emph{Proc.\ DAAD}, May 2022, pp. 1--6.

\bibitem{Farooqi:2023}
M.~N. Farooqi and M.~Ruefenacht, ``Exploring hybrid classical-quantum compute systems through simulation,'' in \emph{IEEE QCE}, vol.~02, 2023, pp. 127--133.

\bibitem{Elsharkawy:2023}
\BIBentryALTinterwordspacing
A.~Elsharkawy, X.-T.~M. To, P.~Seitz \emph{et~al.}, ``Integration of {Quantum} {Accelerators} with {High} {Performance} {Computing} -- {A} {Review} of {Quantum} {Programming} {Tools},'' Sep. 2023, arXiv:2309.06167 [quant-ph]. [Online]. Available: \url{http://arxiv.org/abs/2309.06167}
\BIBentrySTDinterwordspacing

\bibitem{Campbell:2023}
\BIBentryALTinterwordspacing
C.~Campbell, F.~T. Chong, D.~Dahl \emph{et~al.}, ``Superstaq: Deep optimization of quantum programs,'' in \emph{IEEE QCE}, sep 2023, pp. 1020--1032. [Online]. Available: \url{https://doi.ieeecomputersociety.org/10.1109/QCE57702.2023.00116}
\BIBentrySTDinterwordspacing

\bibitem{Hoefler:2015}
\BIBentryALTinterwordspacing
T.~Hoefler and R.~Belli, ``Scientific benchmarking of parallel computing systems: twelve ways to tell the masses when reporting performance results,'' in \emph{Proc.\ Int.\ Conf.\ for HPC, Networking, Storage and Analysis}.\hskip 1em plus 0.5em minus 0.4em\relax ACM, 2015. [Online]. Available: \url{https://doi.org/10.1145/2807591.2807644}
\BIBentrySTDinterwordspacing

\bibitem{Wintersperger:2022}
\BIBentryALTinterwordspacing
K.~Wintersperger, H.~Safi, and W.~Mauerer, ``{QPU}-{System} {Co}-design for {Quantum} {HPC} {Accelerators},'' in \emph{Proc.\ ARCS}, 2022, pp. 100--114. [Online]. Available: \url{https://link.springer.com/10.1007/978-3-031-21867-5_7}
\BIBentrySTDinterwordspacing

\bibitem{safi:23:codesign}
H.~Safi, K.~Wintersperger, and W.~Mauerer, ``Influence of hw-sw-co-design on quantum computing scalability,'' in \emph{IEEE International Conference on Quantum Software}, 2023, pp. 104--115.

\bibitem{Barnes:2008}
\BIBentryALTinterwordspacing
B.~J. Barnes, B.~Rountree, D.~K. Lowenthal \emph{et~al.}, ``A regression-based approach to scalability prediction,'' in \emph{Proceedings of the 22nd Annual International Conference on Supercomputing}, ser. ICS '08.\hskip 1em plus 0.5em minus 0.4em\relax New York, NY, USA: Association for Computing Machinery, 2008, p. 368–377. [Online]. Available: \url{https://doi.org/10.1145/1375527.1375580}
\BIBentrySTDinterwordspacing

\bibitem{Di2012-xf}
S.~Di, D.~Kondo, and W.~Cirne, ``Host load prediction in a google compute cloud with a bayesian model,'' in \emph{2012 International Conference for High Performance Computing, Networking, Storage and Analysis}.\hskip 1em plus 0.5em minus 0.4em\relax IEEE, Nov. 2012.

\bibitem{Calotoiu2013-oe}
A.~Calotoiu, T.~Hoefler, M.~Poke \emph{et~al.}, ``Using automated performance modeling to find scalability bugs in complex codes,'' in \emph{Proceedings of the International Conference on High Performance Computing, Networking, Storage and Analysis}.\hskip 1em plus 0.5em minus 0.4em\relax New York, NY, USA: ACM, Nov. 2013.

\bibitem{Zhang:2017}
\BIBentryALTinterwordspacing
Z.~Jiang, E.~G. Rieffel, and Z.~Wang, ``Near-optimal quantum circuit for grover's unstructured search using a transverse field,'' \emph{Phys.\ Rev.\ A}, vol.~95, Jun 2017. [Online]. Available: \url{https://link.aps.org/doi/10.1103/PhysRevA.95.062317}
\BIBentrySTDinterwordspacing

\bibitem{Wang:2020}
\BIBentryALTinterwordspacing
Z.~Wang, N.~C. Rubin, J.~M. Dominy \emph{et~al.}, ``$xy$ mixers: Analytical and numerical results for the quantum alternating operator ansatz,'' \emph{Phys. Rev. A}, vol. 101, Jan 2020. [Online]. Available: \url{https://link.aps.org/doi/10.1103/PhysRevA.101.012320}
\BIBentrySTDinterwordspacing

\bibitem{Baertschi:2020}
A.~Bärtschi and S.~Eidenbenz, ``Grover {Mixers} for {QAOA}: Shifting complexity from mixer design to state preparation,'' in \emph{Proc.\ IEEE QCE}, 2020, pp. 72--82.

\bibitem{Zhu:2022}
\BIBentryALTinterwordspacing
L.~Zhu, H.~L. Tang, G.~S. Barron \emph{et~al.}, ``Adaptive quantum approximate optimization algorithm for solving combinatorial problems on a quantum computer,'' \emph{Phys. Rev. Res.}, vol.~4, Jul 2022. [Online]. Available: \url{https://link.aps.org/doi/10.1103/PhysRevResearch.4.033029}
\BIBentrySTDinterwordspacing

\bibitem{Tate:2023}
\BIBentryALTinterwordspacing
R.~Tate, J.~Moondra, B.~Gard \emph{et~al.}, ``Warm-{S}tarted {QAOA} with {C}ustom {M}ixers {P}rovably {C}onverges and {C}omputationally {B}eats {G}oemans-{W}illiamson's {M}ax-{C}ut at {L}ow {C}ircuit {D}epths,'' \emph{{Quantum}}, vol.~7, p. 1121, Sep. 2023. [Online]. Available: \url{https://doi.org/10.22331/q-2023-09-26-1121}
\BIBentrySTDinterwordspacing

\bibitem{Vijendran:2024}
V.~Vijendran, A.~Das, D.~Koh \emph{et~al.}, ``An expressive ansatz for low-depth quantum approximate optimisation,'' \emph{Quantum Sci.\ and Tech.}, vol.~9, 02 2024.

\bibitem{Sud:2024}
\BIBentryALTinterwordspacing
J.~Sud, S.~Hadfield, E.~Rieffel \emph{et~al.}, ``Parameter-setting heuristic for the quantum alternating operator ansatz,'' \emph{Phys. Rev. Res.}, vol.~6, May 2024. [Online]. Available: \url{https://link.aps.org/doi/10.1103/PhysRevResearch.6.023171}
\BIBentrySTDinterwordspacing

\bibitem{montanezbarrera:2024}
\BIBentryALTinterwordspacing
J.~A. Montanez-Barrera, D.~Willsch, and K.~Michielsen, ``Transfer learning of optimal qaoa parameters in combinatorial optimization,'' 2024. [Online]. Available: \url{https://arxiv.org/abs/2402.05549}
\BIBentrySTDinterwordspacing

\bibitem{Streif:2020}
\BIBentryALTinterwordspacing
M.~Streif and M.~Leib, ``Training the quantum approximate optimization algorithm without access to a quantum processing unit,'' \emph{Quantum Science and Technology}, vol.~5, May 2020. [Online]. Available: \url{https://iopscience.iop.org/article/10.1088/2058-9565/ab8c2b}
\BIBentrySTDinterwordspacing

\bibitem{noise-model}
\BIBentryALTinterwordspacing
K.~Georgopoulos, C.~Emary, and P.~Zuliani, ``Modeling and simulating the noisy behavior of near-term quantum computers,'' \emph{PRA}, Dec. 2021. [Online]. Available: \url{https://link.aps.org/doi/10.1103/PhysRevA.104.062432}
\BIBentrySTDinterwordspacing

\bibitem{felix}
\BIBentryALTinterwordspacing
F.~Greiwe, T.~Krüger, and W.~Mauerer, ``Effects of imperfections on quantum algorithms: A software engineering perspective,'' in \emph{Proc.\ IEEE International Conference on Quantum Software}, 2023. [Online]. Available: \url{https://doi.org/10.1109/QSW59989.2023.00014}
\BIBentrySTDinterwordspacing

\bibitem{noisy-qaoa}
\BIBentryALTinterwordspacing
C.~Xue, Z.-Y. Chen, Y.-C. Wu \emph{et~al.}, ``Effects of quantum noise on quantum approximate optimization algorithm,'' \emph{Chinese Physics Letters}, vol.~38, mar 2021. [Online]. Available: \url{https://dx.doi.org/10.1088/0256-307X/38/3/030302}
\BIBentrySTDinterwordspacing

\bibitem{noisy-qaoa2}
\BIBentryALTinterwordspacing
J.~Marshall, F.~Wudarski, S.~Hadfield \emph{et~al.}, ``Characterizing local noise in {QAOA} circuits,'' \emph{{IOP} {SciNotes}}, vol.~1, no.~2, aug 2020. [Online]. Available: \url{https://doi.org/10.1088%2F2633-1357%2Fabb0d7}
\BIBentrySTDinterwordspacing

\bibitem{DBLP:journals/corr/BianGB16}
Y.~Bian, A.~Gronskiy, and J.~M. Buhmann, ``Greedy maxcut algorithms and their information content,'' in \emph{IEEE Inf.\ Theory WS}, 2015, pp. 1--5.

\bibitem{vertex-cover-application}
A.~Hossain, E.~Lopez, S.~Halper \emph{et~al.}, ``Automated design of thousands of nonrepetitive parts for engineering stable genetic systems,'' \emph{Nature Biotech.}, 12 2020.

\bibitem{subset-sum-fptas}
\BIBentryALTinterwordspacing
O.~H. Ibarra and C.~E. Kim, ``Fast approximation algorithms for the knapsack and sum of subset problems,'' \emph{J. ACM}, vol.~22, no.~4, oct 1975. [Online]. Available: \url{https://doi.org/10.1145/321906.321909}
\BIBentrySTDinterwordspacing

\bibitem{gw-rounding}
\BIBentryALTinterwordspacing
M.~X. Goemans and D.~P. Williamson, ``Improved approximation algorithms for maximum cut and satisfiability problems using semidefinite programming,'' \emph{J. ACM}, vol.~42, no.~6, p. 1115–1145, nov 1995. [Online]. Available: \url{https://doi.org/10.1145/227683.227684}
\BIBentrySTDinterwordspacing

\bibitem{vertex-cover-approximation}
\BIBentryALTinterwordspacing
G.~Karakostas, ``A better approximation ratio for the vertex cover problem,'' \emph{ACM Trans. Algorithms}, vol.~5, no.~4, nov 2009. [Online]. Available: \url{https://doi.org/10.1145/1597036.1597045}
\BIBentrySTDinterwordspacing

\bibitem{Karloff:2002}
H.~Karloff and U.~Zwick, ``A 7/8-approximation algorithm for {MAX} {3SAT}?'' in \emph{Proceedings 38th Annual Symposium on Foundations of Computer Science}.\hskip 1em plus 0.5em minus 0.4em\relax IEEE Comput. Soc, 2002.

\bibitem{Bazgan:2005}
C.~Bazgan, B.~Escoffier, and V.~T. Paschos, ``\BIBforeignlanguage{en}{Completeness in standard and differential approximation classes: {Poly-(D)APX-} and ({D)PTAS-completeness}},'' \emph{\BIBforeignlanguage{en}{Theor. Comput. Sci.}}, vol. 339, no. 2-3, pp. 272--292, Jun. 2005.

\bibitem{ising}
\BIBentryALTinterwordspacing
A.~Lucas, ``Ising formulations of many {NP} problems,'' \emph{Frontiers in Physics}, vol.~2, 2014. [Online]. Available: \url{http://journal.frontiersin.org/article/10.3389/fphy.2014.00005/abstract}
\BIBentrySTDinterwordspacing

\bibitem{Schoenberger:2023}
M.~Schönberger, I.~Trummer, and W.~Mauerer, ``Quantum optimisation of general join trees,'' in \emph{Proceedings of the International Workshop on Quantum Data Science and Management}, ser. QDSM '23, 08 2023.

\bibitem{Gogeissl:2024}
\BIBentryALTinterwordspacing
M.~Gogeissl, H.~Safi, and W.~Mauerer, ``Quantum data encoding patterns and their consequences,'' in \emph{Proceedings of the 1st Workshop on Quantum Computing and Quantum-Inspired Technology for Data-Intensive Systems and Applications}, ser. Q-Data '24.\hskip 1em plus 0.5em minus 0.4em\relax New York, NY, USA: Association for Computing Machinery, 2024, p. 27–37. [Online]. Available: \url{https://doi.org/10.1145/3665225.3665446}
\BIBentrySTDinterwordspacing

\bibitem{rqaoa2}
\BIBentryALTinterwordspacing
E.~Bae and S.~Lee, ``Recursive {QAOA} outperforms the original {QAOA} for the {MAX}-{CUT} problem on complete graphs,'' \emph{Quantum Information Processing}, vol.~23, no.~3, p.~78, Feb. 2024. [Online]. Available: \url{https://link.springer.com/10.1007/s11128-024-04286-0}
\BIBentrySTDinterwordspacing

\bibitem{max3sat-qubo}
\BIBentryALTinterwordspacing
D.~Ratke. (2021) Qubos for tsp and maximum-3sat. [Online]. Available: \url{https://blog.xa0.de/post/QUBOs-for-TSP-and-Maximum---3SAT/}
\BIBentrySTDinterwordspacing

\bibitem{Krueger:2020}
\BIBentryALTinterwordspacing
T.~Kr\"{u}ger and W.~Mauerer, ``Quantum annealing-based software components: An experimental case study with sat solving,'' in \emph{Proc.\ ICSEW}.\hskip 1em plus 0.5em minus 0.4em\relax New York, NY, USA: ACM, 2020, p. 445–450. [Online]. Available: \url{https://doi.org/10.1145/3387940.3391472}
\BIBentrySTDinterwordspacing

\bibitem{scipy}
P.~Virtanen, R.~Gommers, T.~E. Oliphant \emph{et~al.}, ``{{SciPy} 1.0: Fundamental Algorithms for Scientific Computing in Python},'' \emph{Nature Methods}, vol.~17, pp. 261--272, 2020.

\bibitem{cobyla1}
\BIBentryALTinterwordspacing
M.~J.~D. Powell, ``A {Direct} {Search} {Optimization} {Method} {That} {Models} the {Objective} and {Constraint} {Functions} by {Linear} {Interpolation},'' in \emph{Advances in {Optimization} and {Numerical} {Analysis}}.\hskip 1em plus 0.5em minus 0.4em\relax Springer, 1994. [Online]. Available: \url{http://link.springer.com/10.1007/978-94-015-8330-5_4}
\BIBentrySTDinterwordspacing

\bibitem{list-scheduling1}
R.~L. Graham, ``Bounds for certain multiprocessing anomalies,'' \emph{The Bell System Technical Journal}, vol.~45, no.~9, pp. 1563--1581, 1966.

\bibitem{Cormen:2022}
T.~H. Cormen, C.~E. Leiserson, R.~L. Rivest \emph{et~al.}, \emph{Introduction to algorithms}, fourth edition~ed.\hskip 1em plus 0.5em minus 0.4em\relax Cambridge, Massachusett: The MIT Press, 2022.

\bibitem{noise-model2}
\BIBentryALTinterwordspacing
C.~Blank, D.~K. Park, J.-K.~K. Rhee \emph{et~al.}, ``Quantum classifier with tailored quantum kernel,'' \emph{npj Quantum Information}, vol.~6, no.~1, p.~41, May 2020. [Online]. Available: \url{https://www.nature.com/articles/s41534-020-0272-6}
\BIBentrySTDinterwordspacing

\bibitem{noise-model-evaluation}
\BIBentryALTinterwordspacing
T.~Weber, K.~Borras, K.~Jansen \emph{et~al.}, ``Construction and volumetric benchmarking of quantum computing noise models,'' \emph{Physica Scripta}, vol.~99, no.~6, may 2024. [Online]. Available: \url{https://dx.doi.org/10.1088/1402-4896/ad406c}
\BIBentrySTDinterwordspacing

\bibitem{maschek:25:noise}
\BIBentryALTinterwordspacing
S.~R. Maschek, J.~Schwittalla, M.~Franz \emph{et~al.}, ``\BIBforeignlanguage{en}{Make some noise! measuring noise model quality in real-world quantum software},'' in \emph{\BIBforeignlanguage{en}{Proceedings of the IEEE International Conference on Quantum Software (QSW)}}, 07 2025. [Online]. Available: \url{https://arxiv.org/abs/2506.03636}
\BIBentrySTDinterwordspacing

\bibitem{nielsen-chuang}
M.~A. Nielsen and I.~L. Chuang, \emph{Quantum Computation and Quantum Information}.\hskip 1em plus 0.5em minus 0.4em\relax Cambridge University Press, 2010.

\bibitem{superconducting-qubits}
\BIBentryALTinterwordspacing
P.~Krantz, M.~Kjaergaard, F.~Yan \emph{et~al.}, ``A quantum engineer{\textquotesingle}s guide to superconducting qubits,'' \emph{Applied Physics Reviews}, vol.~6, no.~2, jun 2019. [Online]. Available: \url{https://doi.org/10.1063%2F1.5089550}
\BIBentrySTDinterwordspacing

\bibitem{thelen:24:noisy-qaoa}
\BIBentryALTinterwordspacing
S.~Thelen, H.~Safi, and W.~Mauerer, ``Approximating under the influence of quantum noise and compute power,'' in \emph{Proceedings of WIHPQC@IEEE QCE}, 09 2024. [Online]. Available: \url{http://arxiv.org/abs/2408.02287}
\BIBentrySTDinterwordspacing

\bibitem{Hansen:1993}
P.~Hansen and M.~Zheng, ``\BIBforeignlanguage{en}{Sharp bounds on the order, size, and stability number of graphs},'' \emph{\BIBforeignlanguage{en}{Networks (N. Y.)}}, vol.~23, no.~2, pp. 99--102, Mar. 1993.

\bibitem{Ferrari:2004}
S.~Ferrari and F.~Cribari-Neto, ``Beta regression for modelling rates and proportions,'' \emph{J. Appl. Stat.}, vol.~31, no.~7, pp. 799--815, Aug. 2004.

\bibitem{Box:1964}
G.~E.~P. Box and D.~R. Cox, ``\BIBforeignlanguage{en}{An analysis of transformations},'' \emph{\BIBforeignlanguage{en}{J. R. Stat. Soc. Series B Stat. Methodol.}}, vol.~26, no.~2, pp. 211--243, Jul. 1964.

\bibitem{Georgescu:2014}
I.~M. Georgescu, S.~Ashhab, and F.~Nori, ``\BIBforeignlanguage{en}{Quantum simulation},'' \emph{\BIBforeignlanguage{en}{Rev. Mod. Phys.}}, vol.~86, no.~1, pp. 153--185, Mar. 2014.

\bibitem{Hangleiter:2022}
D.~Hangleiter, J.~Carolan, and K.~P.~Y. Th{\'e}bault, \emph{\BIBforeignlanguage{en}{Analogue quantum simulation}}, 1st~ed., ser. SpringerBriefs in Philosophy.\hskip 1em plus 0.5em minus 0.4em\relax Cham, Switzerland: Springer Nature, Jan. 2022.

\bibitem{Daley:2022}
A.~J. Daley, I.~Bloch, C.~Kokail \emph{et~al.}, ``\BIBforeignlanguage{en}{Practical quantum advantage in quantum simulation},'' \emph{\BIBforeignlanguage{en}{Nature}}, vol. 607, no. 7920, pp. 667--676, Jul. 2022.

\bibitem{Lamata:2018}
L.~Lamata, A.~Parra-Rodriguez, M.~Sanz \emph{et~al.}, ``\BIBforeignlanguage{en}{Digital-analog quantum simulations with superconducting circuits},'' \emph{\BIBforeignlanguage{en}{Adv. Phys. X.}}, vol.~3, no.~1, p. 1457981, Jan. 2018.

\bibitem{qubit-routing1}
\BIBentryALTinterwordspacing
A.~Cowtan, S.~Dilkes, R.~Duncan \emph{et~al.}, ``On the {Qubit} {Routing} {Problem},'' in \emph{Theory of Quantum Computation, Communication, and Cryptography}, 2019. [Online]. Available: \url{https://api.semanticscholar.org/CorpusID:67788345}
\BIBentrySTDinterwordspacing

\bibitem{qubit-routing2}
\BIBentryALTinterwordspacing
G.~Li, Y.~Ding, and Y.~Xie, ``Tackling the {Qubit} {Mapping} {Problem} for {NISQ}-{Era} {Quantum} {Devices},'' in \emph{Proceedings of {ASPLOS XV}}.\hskip 1em plus 0.5em minus 0.4em\relax ACM, Apr. 2019, pp. 1001--1014. [Online]. Available: \url{https://dl.acm.org/doi/10.1145/3297858.3304023}
\BIBentrySTDinterwordspacing

\bibitem{qubit-routing3}
\BIBentryALTinterwordspacing
S.~Niu, A.~Suau, G.~Staffelbach \emph{et~al.}, ``A {Hardware}-{Aware} {Heuristic} for the {Qubit} {Mapping} {Problem} in the {NISQ} {Era},'' \emph{IEEE Transactions on Quantum Engineering}, vol.~1, pp. 1--14, 2020. [Online]. Available: \url{https://ieeexplore.ieee.org/document/9205650/}
\BIBentrySTDinterwordspacing

\bibitem{Hagge:2020}
T.~Hagge, ``Optimal fermionic swap networks for hubbard models,'' 2020.

\bibitem{Weidenfeller:2022}
J.~Weidenfeller, L.~C. Valor, J.~Gacon \emph{et~al.}, ``\BIBforeignlanguage{en}{Scaling of the quantum approximate optimization algorithm on superconducting qubit based hardware},'' \emph{\BIBforeignlanguage{en}{Quantum}}, vol.~6, no. 870, p. 870, Dec. 2022.

\bibitem{Mertens:2003}
\BIBentryALTinterwordspacing
S.~Mertens, ``The easiest hard problem: Number partitioning,'' 2003. [Online]. Available: \url{https://arxiv.org/abs/cond-mat/0310317}
\BIBentrySTDinterwordspacing

\bibitem{Dunning:2018}
I.~Dunning, S.~Gupta, and J.~Silberholz, ``\BIBforeignlanguage{en}{What works best when? a systematic evaluation of heuristics for {Max-Cut} and {QUBO}},'' \emph{\BIBforeignlanguage{en}{INFORMS J. Comput.}}, vol.~30, no.~3, pp. 608--624, Aug. 2018.

\bibitem{McGeoch2023:}
\BIBentryALTinterwordspacing
C.~C. McGeoch and P.~Farr\'{e}, ``Milestones on the quantum utility highway: Quantum annealing case study,'' \emph{ACM TQC}, vol.~5, dec 2023. [Online]. Available: \url{https://doi.org/10.1145/3625307}
\BIBentrySTDinterwordspacing

\end{thebibliography}
\end{document}